\documentclass[twocolumn,showpacs,preprintnumbers,amsmath,amssymb,prl]{revtex4}% Physical Review B

\usepackage{graphicx}% Include figure files
\usepackage{dcolumn}% Align table columns on decimal point
\usepackage{bm}% bold math

\begin{document}

%\preprint{APS/123-QED}

\title{Electron energy relaxation by phonons in the Kondo condensate}

\author{L. J. Taskinen}
\author{I. J. Maasilta}
\affiliation{Nanoscience Center, Department of Physics, P.O. Box 35, FIN-40014 University of Jyv\"{a}skyl\"{a},Finland}
%\author{H. J. H\"{a}kk\"{a}nen}
%\affiliation{Nanoscience Center, Department of Chemistry, P.O. Box 35, FIN-40014 University of %Jyv\"{a}skyl\"{a},Finland}

\date{\today}% It is always \today, today,

\begin{abstract}
 We have used normal metal-insulator-superconductor tunnel junctions as thermometers at sub-Kelvin temperatures to study the electron-phonon (e-p) %%@
interaction in thin Aluminum films doped with Manganese, as a function of Manganese concentration. Mn in Al is known to be a Kondo impurity with %%@
extremely high Kondo temperature $T_K \sim$ 500 K, thus our results probe the e-p coupling in the fully spin compensated, unitary limit. The %%@
temperature dependence of the e-p interaction is consistent with the existing theory for disordered metals, however full theory including the Kondo %%@
effect has not been worked out yet. The strength of the interaction decreases with increasing Manganese concentration, providing a means to improve %%@
sensitivity of detectors and efficiency of solid state coolers.   
\end{abstract}

\maketitle

Recently, there has been a renaissance of the many body physics of magnetic impurities interacting with conduction electrons in metals, known as the %%@
Kondo effect \cite{hewson}. In particular, a lot of understanding has emerged on how the magnetic impurities affect the dephasing \cite{freddy} and %%@
energy relaxation  \cite{huard,anthore} of electrons at low temperatures, by a mechanism where the electron-electron (e-e) interaction is mediated by %%@
the exchange interaction with the magnetic impurities \cite{kim,garst,zarand}. Although this new channel for the e-e interaction is clearly dominant %%@
for dephasing in most impurity-host systems, where the Kondo temperature $T_K$  is much smaller \cite{huard, anthore} or slightly higher %%@
\cite{bauerle} than the electron temperature, it will disappear in the limit where $T << T_K$. In this extreme unitary limit, the impurity spin will %%@
be completely screened by the electrons, will only scatter electrons elastically, and Fermi-liquid theory is know to be valid, if the number of %%@
orbital channels $k$ of the impurity equals twice the spin, $k=2S$ \cite{noz,hewson}.   

In light of the above, we have decided to study an impurity-host system with an extremely large Kondo temperature $T_K \approx$ 500 K, Aluminum doped %%@
with Manganese \cite{hewson}. The expected magnetic ion in this system is Mn$^{2+}$, and if Hund's rule applies, it has five spin aligned d-electrons, %%@
thus $k=2S=5$. As the Kondo impurity mediated mechanism is then suppressed at our experimental sub-Kelvin temperature range, only direct e-e %%@
interaction and electron-phonon (e-p) interaction remain as the main mechanisms for electron energy loss. Here, we concentrate on studying the e-p %%@
interaction, which is typically the dominant energy loss channel for the electron system as a whole. This means it is very critical for the %%@
performance of bolometric detectors and solid state coolers \cite{giazotto}. In fact, AlMn has already been used in novel NIS  bolometers \cite{dan}, %%@
TES microcalorimeters \cite{deiker} and thin-film solid-state refrigerators \cite{clark}.   

In this work, we present the first experimental results on the e-p interaction in the AlMn Kondo-system. No previous data exist in the literature, to %%@
our knowledge. We have measured the temperature dependence and strength of the interaction for samples with varying Mn concentration below 1K, using %%@
NIS tunnel junction thermometry \cite{row}. Results show that e-p interaction weakens significantly below the value for clean Al thin films %%@
\cite{sant}. The strength of the effect depends strongly on the concentration of Mn impurities present.           

The scattering rate between electrons and acoustic phonons via deformation potential depends strongly on temperature, $1/\tau_{e-p} \sim T^m$, where %%@
$m$ can have values ranging between 2-7 depending on whether the sample is a metal or a semiconductor, the type and level of disorder and %%@
dimensionalities of the electron and phonon systems \cite{sergeev1,sergeev2,cle}. For disordered 3D metal films in the limit $ql < 1$, where $q$ is %%@
the wavevector of the dominant thermal phonons and $l$ the electron mean free path, theory predicts \cite{pipp,schmid,alt,reiz} that if the scatterers %%@
are vibrating with the phonons, interference terms lead to suppression of e-p relaxation. Then $m=4$ (in contrast to $m=3$ for pure samples), and the %%@
net power flowing from hot electrons of volume $\cal V$ at a temperature $T_e$ to phonons at $T_p$ is 
\begin{equation} 
P_{e-p} = \Sigma{\cal V} (T_e^n-T_p^n),
\label{eq:ep}
\end{equation}
where $n=m+2=6$ and $\Sigma$ is a sample parameter. This suppression has recently been observed in Cu and Au noble metal films \cite{jenni}, where the %%@
simple theory is expected to be correct, in addition to earlier results on Ti and Hf \cite{gersh}. The effects of the more complex Fermi surface of Al %%@
and the inclusion of the Mn Kondo ions are presently not known theoretically. 

\begin{figure}[h]
\includegraphics[width=0.9\linewidth]{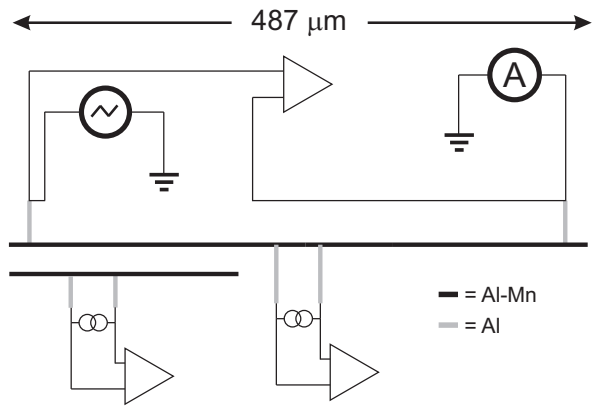}%[width=0.9\linewidth]
\caption{\label{fig:skema}Schematic of the samples and the measuring circuit.}
\end{figure}

A Schematic of the samples used to measure the e-p interaction is shown in Fig. \ref{fig:skema}. We use the hot-electron technique \cite{roukes} to %%@
measure the e-p interaction by overheating the electrons by Joule heat power $P$ and measuring the resulting temperature $T_e$. All the measured %%@
samples had two long ($L\sim 0.5$ mm), electrically isolated Al-Mn normal metal wires separated by 2 $\mu$m. Since $L >> L_{e-e}$, the %%@
electron-electron scattering length, $T_e$ is well defined without complications from non-equilibrium \cite{anthore}. The upper wire was heated by %%@
applying a voltage across a pair of superconducting Al leads in direct metallic contact to AlMn, forming SN junctions. These junctions provide %%@
excellent electrical, but very poor thermal conductance due to Andreev reflection, as the junctions are biased within the superconducting gap $\Delta$ %%@
of Al in all the data shown here. Thus, due to the lack of outdiffusion of electrons, the input heat is distributed uniformly in the interior of the %%@
wire \cite{diffu,nima}.  We measure the electron temperature in the middle of the wire, where two additional Al leads form a NIS tunnel junction pair %%@
(SINIS), as a function of input the Joule power $P=IV$, measured in four probe configuration. The purpose of the lower Al-Mn wire, with an additional %%@
SINIS thermometer on it, is to give an upper limit to the local $T_p$, as the e-p power flow (Eq. \ref{eq:ep}) depends on both $T_e$ and $T_p$. As our %%@
sample volumes are large, the contribution of thermal photon emission \cite{ssc} is estimated to be insignificant compared with phonon emission.

Several e-p samples were fabricated in three different Mn concentrations (S1,S2,S3), on nitridized silicon chips using electron beam lithography and %%@
three-angle shadow evaporation of Al and Al-Mn \cite{source}. Four probe samples with similar Al-Mn wire dimensions to the e-p samples were also %%@
fabricated for accurate resistivity measurements between room temperature and 4K (samples RS1, RS2 and RS3). Evaporation was done from zero angle %%@
using the same evaporation parameters (growth rate, new crucible every time and same amount of source material) as for the e-p samples. 
The oxide layer forming the tunnel barrier was produced by thermal oxidation of Al, resulting in tunneling resistances $R_T~\sim$ 6-15 k$\Omega$.   %%@
Widths of the Al-Mn wires varied from 300 to 500 nm, thicknesses from 50 to 60 nm, and length was always $\sim 487\mu$m. The small dimensions were %%@
measured accurately for each sample with an AFM.  

\begin{figure}[h]
\includegraphics[width=1\linewidth]{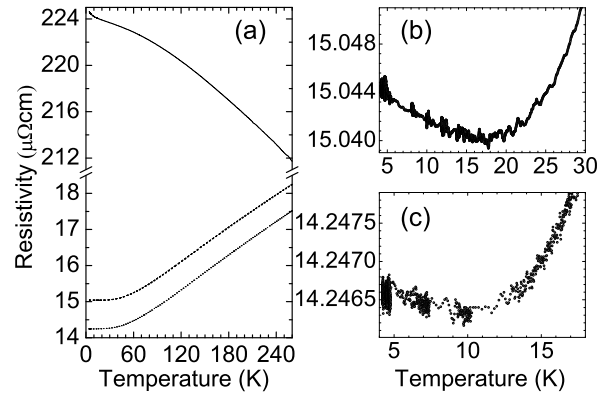}%[width=0.9\linewidth]
\caption{\label{fig:roo}(a) Temperature dependence of the resistivity: continuous line sample RS1, dashed RS2 and dotted RS3. Blow-up of the low %%@
temperature range of the data for RS2 and RS3 are shown in (b) and (c), respectively.}
\end{figure}

The composition of the evaporated AlMn films were determined using elastic recoil detection analysis \cite{erda}. Results for samples RS1, RS2 and RS3 %%@
show atomic Mn concentrations  $c_i$ of ($13.6\pm 1.0$) \%, ($1.3\pm 0.2$) \% and ($0.54\pm 0.10$)\%, respectively. These can be compared with the %%@
nominal source concentrations of 2 \%, 0.65 \% and 0.3 \% for RS1-RS3, respectively. In evaporation, the Mn concentration increases due to the larger %%@
vapor pressure of Mn compared to that of Al.

\begin{figure}[h]
\includegraphics[width=0.9\linewidth]{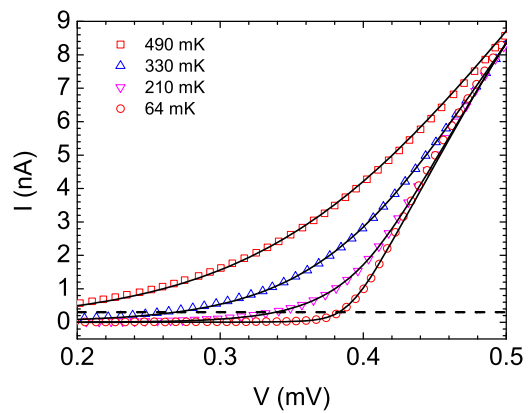}% Here is how to import EPS art
\caption{\label{fig:IV}Measured (symbols) current-voltage characteristics of a SINIS structure (nominally 0.65 \% Mn) at different refrigerator %%@
temperatures between 60 - 500 mK. Solid lines are calculated from BCS theory. At $T=64$ mK, the thermal model described in Ref. \cite{clark} was used.  %%@
The value of the only fitting parameter, the fraction of the power dissipated in Al returning to AlMn via recombination phonons, is $\beta=0.3$. %%@
Dashed line corresponds to the bias current used in the e-p measurements.}
\end{figure}

Data from resistivity measurements is shown in Fig. \ref{fig:roo}. Samples RS2 and RS3 exhibit a typical metallic resistivity, where the phonon %%@
contribution dominates at high temperatures (region where $d\rho/dT > 0$), whereas below 30 K the impurity resistivity dominates. However, if one %%@
looks at the low temperature region more carefully [Figs. \ref{fig:roo} (b) and (c)], clear minima for both samples can be seen. This is a signature %%@
of the Kondo effect. In sample RS1 with the highest Mn concentration, the contribution of the Kondo resistivity is so large that it completely %%@
overwhelms the phonon contribution in the whole temperature range all the way up to room temperature ($d\rho/dT < 0$). The shape of the $\rho$ vs. $T$ %%@
curve agrees qualitatively with the Kondo theory in the strong coupling limit $T < T_K$. Quantitative comparison is difficult, as the temperature %%@
range is such that only numerical renormalization group calculations are valid \cite{hewson}. The residual resistivity at 4.2 K is approximately %%@
linear in $c_i$, as expected for the unitary limit \cite{hewson}.

\begin{figure}[h]
\includegraphics[width=0.9\linewidth]{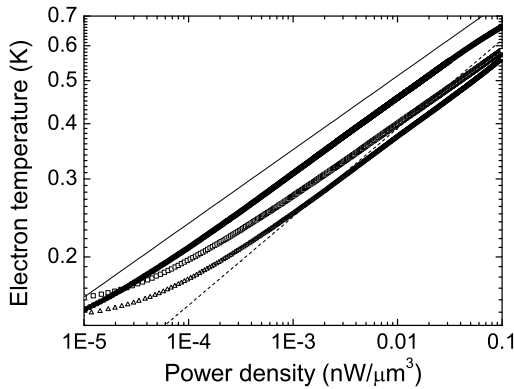}% Here is how to import EPS art
\caption{\label{fig:data} Measured electron temperatures of the AlMn e-p samples  vs. applied power per unit volume. Circles, squares and triangles %%@
denote data from samples S1, S2 and S3, respectively. Continuous line is a guide for the eye $\propto T^{1/6}$. Dashed line corresponds to Al data in %%@
the pure limit, $n=5$, with $\Sigma=1.1\times10^9$/K$^5$m$^3$ calculated for sample 1 in Ref. \cite{sant}, using a value for Sommerfeld constant %%@
$\gamma=135$J/K$^2$m$^3$.}
\end{figure}

Having established the relevance of Kondo physics in our samples, we discuss next the e-p measurements, performed in a dilution refrigerator at base %%@
temperature $T= 60$ mK. As AlMn NIS junctions have not been used widely in thermometry before, we first have to ensure that they work as expected. %%@
Fig. \ref{fig:IV} shows a typical current-voltage characteristics of a AlMn/AlOx/Al SINIS junction, with the refrigerator temperature varying between %%@
60-500 mK. The usual suppression of the current at small bias voltages due to the superconducting gap ($2\Delta/e \sim 0.4$ mV) is seen. As $T$ is %%@
increased, the gap feature is smeared and more current flows. Thus, by current biasing the the junction and measuring its voltage, very sensitive %%@
thermometry is possible. Fig. \ref{fig:IV} also shows theoretical modeling based on BCS theory and a thermal model introduced in Ref. \cite{clark}. %%@
This model takes into account the current and heat flows across the junctions calculated from the exact BCS formulas, in addition to cooling by %%@
phonons, back emission of heat from the superconductor, and dissipation due to imperfections of the barrier. We found that only at 60 mK was the full %%@
thermal model necessary for a good fit (with one free parameter), at all other temperatures a simple BCS theory with a constant $T$ was sufficient to %%@
model the data. This was checked for all Mn concentrations used. Thus, the SINIS junctions could be calibrated using a calibrated RuO thermometer on %%@
the sample stage,  taking into account noise overheating at the lowest temperatures, as discussed in Refs. \cite{statsol,jenni}. 

\begin{table}
\caption{\label{tab:table}Parameters of the e-p samples. the mean free path $l$ was calculated from the resistivity at $T=60$ mK using the Drude %%@
formula, $\Sigma$ was obtained from the fits to the data in Fig. \ref{fig:data}, $r=(1/\tau_{AlMn})/(1/\tau_{Al})=6\Sigma T/(5 \Sigma_{Al})$. }
\begin{ruledtabular}
\begin{tabular}{ccccc}
Sample & Source Mn \% &$l$ (nm) & $\Sigma$ ($10^9$W/K$^6$m$^3$) & $r$(0.1 K)\\
\hline
S1 & 2    & 0.2 & 1.1 & 0.12 \\
S2 & 0.65 & 1.9 & 2.5 & 0.27 \\
S3 & 0.3  & 3.2 & 4.2 & 0.46 \\
\end{tabular}
\end{ruledtabular}
\end{table}

% In the model for a SINIS at thermal equilibrium
%\begin{equation} 
%\sum{_1,2}P_{cool}+\beta (P_{cool}+IV)+P_{s}-P_{e-p} = 0,
%\label{eq:Peq}
%\end{equation}
%where $P_{cool}$ is the cooling power of the SINIS, $I$ and $V$ are current and voltage of the SINIS, $\beta 
%(P_{cool}+IV)$ is the fraction of the heat put to the superconductor flowing back to the normal metal,  $P_{s}=V^2/R_s$ is dissipation due to %%@
%imperfect tunnel barrier, and $R_s$ is the zero bias resistance. $P_{e-p}$ is calculated from Eq. \ref{eq:ep}. In the calculation we used $\Sigma$ = %%@
%2.5$\times$10$^9$ W/K$^6$m$^3$ and $\Omega$=1.1$\times$10$^-17$m$^2$ got from e-p and AFM measurements. For $\beta$, which is the only fitting %%@
%parameter in this case, we got $\beta$=0.3. 
%Current-voltage characteristics of all the SINIS thermometers used in this work were measured at least at the refrigerators base temperature $\sim$60 %%@
%mK and found to be BCS like. 

%Thermometers were calibrated by slowly changing the cryostat temperature and comparing SINIS voltage to a calibrated RuO thermometer attached to the %%@
%sample stage.  Measurement and SINIS calibration procedures are described in more detail elsewhere \cite{statsol, jenni}.

\begin{figure}[h]
\includegraphics[width=0.9\linewidth]{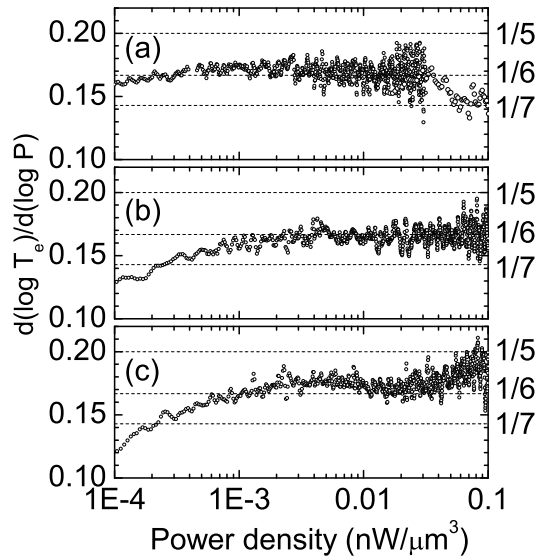}% Here is how to import EPS art
\caption{\label{fig:logder} Numerical logarithmic derivatives of the measured data in Fig. \ref{fig:data} for samples S1 (a), S2 (b) and S3 (c).}
\end{figure}

 The data for samples S1, S2 and S3 used to obtain the strength of the e-p interaction is shown in Fig. \ref{fig:data}.   We have plotted the electron %%@
temperatures $T_e$  (measured with SINIS thermometers) as a function of the input power per sample volume applied to the electron gas, in log-log %%@
scale. The temperatures of the unheated wires were found to be low enough, so that the condition $T_e^n \gg T_p^n$ holds, and the data can be analyzed %%@
without the direct knowledge of $T_p$. In this case, the slope of the measured $T_e$ vs. $P$ gives directly the value $1/n$, and the constant %%@
multiplier the value $\Sigma$. All samples seem to be consistent with $n=6$, agreeing with similar Cu and Au samples \cite{jenni}. The saturation of %%@
$T_e$ at low $P$ is due to external noise heating \cite{jenni}. Values of $\Sigma$ obtained from fits fixing $n=6$ are shown in Table \ref{tab:table}. %%@
Clearly, the e-p coupling weakens significantly with increasing Mn concentration. Fig. \ref{fig:data} also shows a comparison with the clean limit for %%@
Al films, using $n=5$ and $\Sigma_{Al} = 1.1\times 10^{9}$W/K${^5}$m${^3}$, obtained from Ref. \cite{sant}. Comparing $\Sigma_{Al}$ with our results, %%@
we see that the energy relaxation rate $1/\tau$ is suppressed by an order of magnitude at 100 mK for the highest Mn concentration, see Table %%@
\ref{tab:table}. Comparison with the data measured for Cu and Au wires of approximately the same film thickness using the same technique \cite{jenni} %%@
gives temperature independent suppression factors ranging from 0.02 (AlMn 2\% vs Au) to 0.3 (AlMn 0.3 \% vs Cu).    
Finally, to obtain a detailed picture of the temperature dependence, we plot the numerical derivatives $d(\log T_e)/d(\log P)$ of the data in Fig. %%@
\ref{fig:logder}. It is clear that $P \sim T^6$ describes the data well over a large range of $P$ for all samples. The deviations at highest heating %%@
powers for samples S1 and S3 are presently not understood.   

% In calculating $\Sigma$ we have used Sommerfeld constant for aluminum $\gamma=135$J/K$^2$m$^3$ and $\Sigma \approx 0.524\alpha\gamma$ \cite{wells}, %%@
%where $\alpha$ is the constant in e-p scattering rate equation, $\tau_{e-p}^{-1} = \alpha T^5$, and  deduced from Ref. \cite{sant} for the sample 1 %%@
%therein. 

In conclusion, we have obtained clear evidence for suppression of phonon emission from electrons in the extreme Kondo compensated limit $T << T_K$ in %%@
Aluminum doped with Manganese. As Kondo impurity mediated e-e scattering is expected to be suppressed in the unitary limit (other possible Kondo %%@
impurities such as Cr and Fe, also have a high $T_K> 300$ K in Al), pure e-e scattering should dominate the dephasing at sub-Kelvin temperatures in %%@
this material. The observed suppression of e-p scattering also leads to an extension of the temperature range of e-e dephasing, therefore AlMn is a %%@
good candidate material to study the e-e interaction. For hot-electron bolometers, a weak e-p coupling leads to increased sensitivity. The expected %%@
improvement of the noise equivalent power (NEP $\sim \sqrt{\Sigma}$) for the high concentration AlMn is a factor $\sim 4$ over Cu ($\sim 7$ over Au). %%@
Weak e-p coupling is also benefitial for solid state electron coolers.  

\begin{acknowledgments}
We acknowledge Tarmo Suppula for help in sample fabrication and Timo Sajavaara and Kenichiro Mizohata for the Mn concentration analysis. We are also %%@
grateful to A. C. Hewson and D. E. Prober for valuable discussions. This work was supported by the Academy of Finland under contracts No. 105258 and %%@
205476.  
\end{acknowledgments}

%\begin{thebibliography

%\end{thebibliography}
%\bibliography{almn}% Produces the bibliography via BibTeX.

\begin{thebibliography}{31}
\bibitem[{dif()}]{diffu}
\bibinfo{note}{Electron temperature is nonuniform within the distance
  $L_{e-p}=\sqrt{D\tau_{e-p}}$ away from the NS contacts. As $L >> L_{e-p}$
  here, no complications arise due to end effects, unlike in an earlier initial
  work \cite{nima}.}
\end{thebibliography}

\end{document}